\documentclass[12pt,preprint]{aastex}

\usepackage{etoolbox}

\shorttitle{A Comprehensive Study of Atomic Hydrogen Adsorbing to Amorphous Water-Ice}
\shortauthors{Dupuy et al.}

\begin{document}


\title{A Comprehensive Study of Hydrogen Adsorbing to Amorphous Water-Ice: \\ \textit{Defining Adsorption in Classical Molecular Dynamics}}


\author{John L. Dupuy, Steven P. Lewis, and P. C. Stancil}
\affil{Department of Physics and Astronomy and the Center for Simulational Physics, \\ University of Georgia,
    Athens, GA 30602}



\begin{abstract}
Gas-grain and gas-phase reactions dominate the formation of molecules in the interstellar medium (ISM). Gas-grain reactions require a substrate (e.g. a dust or ice grain) on which the reaction is able to occur. The formation of molecular hydrogen (H$_2$) in the ISM is the prototypical example of a gas-grain reaction. In these reactions, an atom of hydrogen will strike a surface, stick to it, and diffuse across it. When it encounters another adsorbed hydrogen atom, the two can react to form molecular hydrogen and then be ejected from the surface by the energy released in the reaction. We perform in-depth classical molecular dynamics (MD) simulations of hydrogen atoms interacting with an amorphous water-ice surface. This study focuses on the first step in the formation process; the sticking of the hydrogen atom to the substrate. We find that careful attention must be paid in dealing with the ambiguities in defining a sticking event. {The technical definition of a sticking event will affect the computed sticking probabilities and coefficients. Here, using our new definition of a sticking event,} we report sticking probabilities and sticking coefficients for nine different incident kinetic energies of hydrogen atoms [5 K - 400 K] across seven different temperatures of dust grains [10 K - 70 K]. {We find that probabilities and coefficients vary both as a function of grain temperature  and incident kinetic energy over the range of 0.99 - 0.22. }
\end{abstract}


\keywords{ dust, extinction --- ISM: abundances --- molecular processes} 



\section{Introduction}

Gas-phase and gas-grain reactions are considered the two main paths by which molecules form in the Interstellar Medium (ISM) \cite[]{alh07}. {As of February 2016, almost 200 molecular species\footnote{cf. https://www.astro.uni-koeln.de/cdms/molecules} have been observed in the ISM and circumstellar shells}; a number that will only increase as observational techniques improve. It is well known that the observed abundance of molecular hydrogen (H$_2$) cannot be accounted for by gas-phase reactions \cite[]{per05}. Thus, gas-grain reactions are invoked to account for the amount of H$_2$ observed in the ISM. 

The main regions of focus for H$_2$ formation are molecular clouds (MCs) and the cold neutral medium (CNM). These regions have temperatures that range from 10 K to 100 K, and make up a bit under half of the ISM by mass \cite[]{tie05}. The interplay of dense MCs with the CNM allows for the formation of H$_2$ and plays a crucial role in star formation.

There are four main steps in the typical gas-grain reaction. The reactants must first adsorb to the interstellar dust surface; they then must diffuse across the substrate. When reactants encounter each other, they must react to form a molecular species. Finally, the newly formed molecule must desorb from the dust surface.

{In dense clouds, }these interstellar grains/substrates\footnote{In the following discussion, we use the terms ``grain" and ``substrate" interchangeably.} consist of silicate or carbonaceous cores covered by a thin layer of ice about 0.1 $\mu$m thick \cite[]{hol09}. The icy layers are believed to predominantly be composed of H$_2$O; although ices such as CO, CO$_2$, NH$_3$, and CH$_4$ certainly occur as well {\cite[]{alh07,boo15}}. The H$_2$O ice is generally regarded to be amorphous in structure {\cite[]{hag81,boo15}}.

The formation of H$_2$ on dust grain surfaces is thought to occur through three mechanisms. The first, dubbed the Eley-Rideal (ER) mechanism, involves an impinging hydrogen atom directly hitting an already adsorbed hydrogen atom on the icy surface \cite[]{ele41}. In the Langmuir-Hinshelwood (LH) mechanism, two hydrogen atoms already adsorbed to the grain surface diffuse some distance until they meet one other and react to form H$_2$ \cite[]{hin30,lan22}. {The distinction between these two mechanisms lies in the thermalization of the atom with respect to the grain. In the ER mechanism, the adsorbed atom has thermalized to the grain, and is directly hit by an impinging atom that has not yet thermalized. By contrast, in the LH mechanism, both atoms have already thermalized with the grain. } The final mechanism, known as the hot-atom mechanism, involves a hydrogen atom with a high kinetic energy striking the substrate and thermally hopping along the substrate surface until it strikes another already adsorbed hydrogen atom and forms H$_2$ \cite[]{har81}.

The common feature of these three mechanisms is the adsorption of at least one of the hydrogen atoms to the icy substrate. A key quantity characterizing the adsorption properties\footnote{Also called a sticking event in the following discussion.} of a given gas species on a given substrate is the probability that the gas molecule sticks to the substrate when the two collide. In computational studies, the sticking probability can be deduced as the fraction of incident gas molecules that stick to the substrate, in the limit of a large number of trials. The sticking probability for a general species is 
\begin{equation}
P(T_{\rm d},E_{\rm i}) = \frac{N_{\rm stick}}{N_{\rm tot}}, 
\end{equation}  
where $N_{\rm stick}$ is the number of sticking events and $N_{\rm tot}$ is the number of trials. The sticking probability depends on the temperature of the dust or ice grain ($T_{\rm d}$) and the incident kinetic energy ($E_{\rm i}$) of the colliding species when the species and surface are very far apart.  

In many experimental and observational situations, one doesn't have access to the incident kinetic energy of the gaseous species, but rather the temperature of the gas ($T_{\rm i}$). The quantity characterizing the sticking in this case is called the sticking coefficient, $S(T_{\rm i},T_{\rm d})$, a function of both gas and grain temperatures. The sticking coefficient can be calculated by averaging the sticking probability weighted by a Maxwellian velocity distribution of the desired gas temperature. Hollenbach \& Salpeter (1970) derived a theoretical form for this sticking coefficient: 
\begin{equation}
S(T_{\rm i},T_{\rm d}) = \frac{\int_0^\infty\epsilon e^{-\gamma \epsilon}P(T_{\rm d},E_{\rm i}) d\epsilon}{\int_0^\infty\epsilon e^{-\gamma \epsilon} d\epsilon} ,  
\end{equation} 
where $\gamma = E_{\rm c} / k T_{\rm i} $, $\epsilon = E_{\rm i}/E_{\rm c}$,  and $E_{\rm c} = \Omega(D \Delta E_{\rm s})^{1/2}$. The variable $D$ is the well depth of the adsorption potential, $\Delta E_{\rm s}$ is the energy transferred in a single collision, and $\Omega^2$ is close to unity for Lambert's law and equals two in the case of isotropic scattering \cite[]{vee14}.  
In astrophysical models, this equation is often approximated with 
\begin{equation}
S(T_{\rm i},T_{\rm d}) \approx \frac{(\gamma^2 + 0.8\gamma^3)}{(1+2.4\gamma+\gamma^2+0.8\gamma^3)},
\end{equation}
where $\gamma$ is dependent on the incident species and the type of substrate involved \cite[]{vee14}. Some experiments (e.g. Manico et al. 2001) suggest that Eqn. (3) underestimates the sticking coefficient. The sticking coefficient is of great importance to astrophysical models, as it is a crucial component for the rate equation of various grain-mediated processes (including H$_2$ formation) in the ISM. 

The goal of this work is therefore to obtain accurate sticking coefficients of atomic hydrogen sticking on amorphous water-ice for astrophysical modeling purposes.  There have been many such studies, experimental and theoretical, with the same goal, {though H$_2$ is primarily formed on warm amorphous carbon and silicate grains \cite[]{iqb12,leb12}}. To our knowledge, the first simulational study was performed by Buch \& Zhang (1991), who studied the sticking of H and D atoms on an amorphous cluster of 115 water molecules using classical molecular dynamics (MD). They found sticking probabilities and coefficients on 10 K ice for both H and D atoms. A group of studies investigated the sticking and mobility of H atoms on amorphous water-ice slabs consisting of 1000 water molecules using classical MD \cite[]{mas97,tak99}. They performed simulations on 10 K and 70 K ice and showed that the sticking probability cannot be assumed to be equal to unity for the range of gas temperatures encountered in the ISM. However, due to the computational restraints of that era, not many simulations were performed for each incident kinetic energy of the hydrogen atom.   

Al-Halabi et al. (2002, 2007) also performed classical MD simulations on both crystalline and amorphous water-ice substrates. Their simulations confirmed that sticking depends on the incident kinetic energy of the H-atom and the temperature of the ice surface by carrying out a large number of trials. Veeraghattam et al. (2014) performed large numbers of classical MD simulations on both 10 K and 70 K amorphous water-ice slabs consisting of 1000 water molecules. They computed sticking probabilities and coefficients for their simulations that are reproduced and expanded upon in the current work. Experimentally, Manico et al. (2003), Hornekaer et al. (2003), and Watanabe et al. (2010) all observed H$_2$ formation on lab-generated substrates and measured the recombination energy.             

In this study we perform an extensive analysis of atomic hydrogen sticking to amorphous water-ice surfaces for a wide range of grain temperatures (10 K - 70 K) and for a wide range of atomic hydrogen kinetic energies (5 K - 400 K). Therefore a key goal is to fill in the large gaps in predicted sticking coefficients from previous studies to provide richer, more in-depth information for astrophysical modelers. Here we also pay particularly close attention to the definition of what constitutes a sticking event. The previous simulational studies mentioned above do not agree on a objective definition for a sticking event, and we have found that different criteria can lead to different results. The appendix provides additional detail on the subject. 

\section{Computational Methodology}
Using classical molecular dynamics (MD), we study the interaction of a hydrogen atom with an amorphous water-ice slab \cite[]{all87}. It is reasonable to question whether classical dynamics is appropriate for describing hydrogen. Studies of hydrogen in other contexts have shown that quantum dynamics is preferable for hydrogen, but at a very high computational cost. In the meantime, classical trajectory calculations have been shown to prove useful in describing physical processes involving hydrogen on surfaces \cite[]{and06}. But, as Andersson et al. (2006) mention, quantum effects can be tricky to pin down and could be important as the kinetic energy of the hydrogen atom decreases. Nevertheless, at this stage the vastly more complicated and expensive quantum calculations seem unwarranted. 

The present simulations require two pairwise potentials: (1) H$_2$O-H$_2$O and (2) H-H$_2$O. These potentials are differentiated to define a force field, which is then used to numerically integrate the coupled, classical equations of motion for the many-particle system using the leapfrog algorithm at a fixed time-step of $\Delta t = 1$ fs. In non-spherical systems containing rigid molecules, like H$_2$O, the rotational equations of motion must also be solved \cite[]{vee14}.   

The H$_2$O-H$_2$O potential interaction is used to simulate the water-ice substrate while the H-H$_2$O potential is used to model the hydrogen interaction with that substrate. These potentials are derived from quantum chemical calculations for a single pair of particles. Our system is treated either as a canonical ensemble (NVE) or as a isobaric isothermal ensemble (NVT), depending on the stage of the simulation. 

\subsection{Slab Creation}
We consider a system of 1000 water molecules to model the amorphous water-ice surface as a slab of finite thickness. The water molecules are treated as rigid and exist within a $40 $ \AA $\times 40$ \AA $\times 40$ \AA $\:$ simulation box with periodic boundary conditions in the $x,y,$ and $z$ directions. The interaction potential for these water molecules is a TIP4P/Ice configuration \cite[]{aba05}. This TIP4P/Ice model is a slight modification of the well-known transferable intermolecular potential model (TIP4P) \cite[]{jor82}. The basic form consists of a Lennard-Jones (LJ; Lennard-Jones \& Devonshire 1936) term added to a Coulombic term: 
\begin{equation}
V_{\textnormal{H}_2 \textnormal{O} - \textnormal{H}_2 \textnormal{O}} = \sum_{i,j} \frac{q_i q_j e^2}{r_{ij}} + 4\epsilon \left[\left(\frac{\sigma}{r}\right)^{12} - \left(\frac{\sigma}{r}\right)^{6}\right].
\end{equation}

The LJ term applies to a site that is on the oxygen atom but in the Coulombic term, each H$_2$O molecule is modeled by three charge sites: a positive charge on each hydrogen atom and a negative charge on the H-O-H molecular angle bisector, 0.1577 \AA $\:$ from the oxygen atom. The variable $r$ is the distance between two oxygen atoms and $r_{ij}$ is the distance between charge sites on two different H$_2$O molecules, where $i$ and $j$ each run over the three charge sites of the two molecules. The LJ constants have values of $\sigma=3.166$ \AA $\:$ and $\epsilon = 0.211$ kcal mol$^{-1}$. The charge on each hydrogen site is $q_{\rm H} = 0.5897 e$, where $e$ is the electron charge; the charge on the H-O-H molecular angle bisector is $-2q_{\rm H}$. The H-O-H angle has a fixed value of 104.52$^{\circ}$, $r_{\rm OH} = 0.9572 $ \AA $\;$ which is the fixed length of the O-H bond.  

We use the potential above to generate the amorphous ice slab from a quenched MD simulation. Within our simulation box, we restrict the initial molecular position in the slab to a $40 $ \AA $\times 40$ \AA $\times 20$ \AA $\:$ region. The restriction on the $z$ coordinate ensures that we generate a slab rather than a cubic arrangement of water molecules. We give each molecule a random initial position and velocity within this region and the system is then numerically solved. A 10 \AA $\:$ cutoff was used for the Coulombic interaction and Ewald summation was used for long-range electrostatics. 

Quenching the slab occurs as follows: initially, the system evolves for 150 ps in NVT at a temperature of 300 K using velocity rescaling to set the temperature. The system then evolves in NVE for 50 ps, after which the system is cooled to 200 K and run again in NVT for 150 ps, followed by another 50 ps NVE run. The temperature is then lowered to 100 K. The process is repeated at 100 K before we cool the slab to 70 K. {The reason it is desirable to perform a simulation in NVT is to ensure that the temperature of the amorphous ice slab, which is only a small component of an actual interstellar ice mantle, stays approximately constant. As a result of the NVT simulation, the substrate will settle into an energetically favorable configuration. The system is then run in NVE so that the substrate can be fixed in that energy configuration.}

To generate slabs of other temperatures we sequentially cool this 70 K slab in 10 K intervals, i.e. from 70 K to 60 K to ... to 10 K. The system runs in NVE for 10 ps at each interval and velocity rescaling occurs every 5 ps. A restart file is output at the end of each interval, containing the positions and velocities of the water molecules to be used as the initial conditions of the next cooling interval. We tested this method by generating a 10 K slab from scratch and comparing calculated sticking probabilities on that slab to sticking probabilities on a 10 K slab made from the 70 K slab. The sticking probabilities we calculated were in very good agreement, and we determined that this method is sufficient for slab generation at a variety of temperatures. Note that in this method of making the 10 K slab, we also make slabs of the other temperatures studied along the way. These slabs are then used for hydrogen atom trajectory calculations. 

\subsection{Hydrogen Atom Trajectory Calculation}
 
A single hydrogen atom trajectory\footnote{Also called a ``run" in the following discussion.} is initialized as follows. First, we assign the hydrogen atom random initial $x$,$y$ coordinates and a $z$ coordinate fixed at 18 \AA, about 8 \AA $\;$ above the ice surface. Because the ice is amorphous the surface of the slab is ill defined, and therefore we use an average surface for referencing H-atom/surface distances. Random initial velocity components that correspond to a particular kinetic energy are assigned as well. The initial velocity in the $z$-direction must be negative so that the hydrogen atom is launched towards the slab. After the initialization, the trajectory is then integrated numerically using the leapfrog algorithm. 

The interaction between the hydrogen atom and the water molecules of the substrate is described by the potential derived by Zhang et al. (1991), which has a form given by
\begin{equation}
V_{\textnormal{H} - \textnormal{H}_2 \textnormal{O}} = \sum_{l,m} 4\epsilon_{l,m}  \left[\left(\frac{\sigma_{l,m}}{r}\right)^{12} - \left(\frac{\sigma_{l,m}}{r}\right)^{6}\right]Y_{l,m}(\theta,\phi).
\end{equation}
This potential was developed using open-shell Hartee Fock and fourth-order M\o ller-Plesset perturbation theory. The \textit{ab initio} data were then fit to a LJ form given by equation (5). The variable $r$ represents the distance between the impinging hydrogen atom and the oxygen atom. The $\theta$ and $\phi$ coordinates are the spherical angles of the incident hydrogen atom with respect to the water molecule. The origin is at the oxygen atom and the polar axis coincides with the molecular symmetry axis. For a water molecule, the spherical coordinates of the two hydrogen atoms are given by $r_{\rm OH} = 0.9572$ \AA, $\theta' = 127.74^{\circ}$, and $\phi' = \pm 90^{\circ}$. The $\sigma_{l,m}$ and $\epsilon_{l,m}$ parameters for this potential are given in Table 1.

The simulation time for a single run is 10 ps and the time-step is 1 fs. The simulation time was selected in an attempt to avoid ambiguity in determining the sticking or scattering status of the hydrogen atom. Periodic boundary conditions were used for the $x$ and $y$ coordinates (parallel to the slab surface), but the simulation box was non-periodic in the $z$-coordinate (normal to the surface). Thus the hydrogen atom leaves the simulation box -- and the simulation terminates -- if the $z$-coordinate of the atom ever exceeds 20 \AA. During the simulation, the system runs in NVE, but the velocity of the substrate molecules are rescaled every 5 ps to ensure that the slab temperature remains (approximately) constant. 

For nine incident kinetic energies of hydrogen atoms ranging from $E_{\rm H}/k_{\rm B} = $5 K - 400 K, we performed 300 runs apiece for each of the 7 substrate temperatures. Thus, simulation data for a single substrate temperature comprises of 2700 runs, for a total of 18,900 runs over the 7 substrate temperatures. We chose to perform 300 runs for each H-atom to obtain statistically significant values for the sticking probability. For each of the 18,900 runs, the sticking or scattering status was determined. We defined a sticking event to be a trajectory whose final $z$ coordinate ($Z_{\rm f}$) was below 13 \AA $\:$ and whose final total energy ($E_{\rm f}$) was below -100 K. The physical motivation for this definition is discussed in the appendix. To get an estimate of error for each batch of runs we randomly selected 50 runs out of the 300 and calculated the sticking probability. We repeated this random sampling 25 times and computed the standard deviation of the probabilities to represent the error of our computations. 

\section{Results \& Discussion}     

MC and CNM regions of the ISM contain dust grains covered with ice which have temperatures that range from 10 K to 100 K. We have performed simulations on 7 different ice substrate temperatures (10 K - 70 K) for 9 different incident kinetic energies (5 K - 400 K) of the H-atom. This wide selection of slab temperatures provides a thorough coverage of the temperature ranges of interest. All of our sticking probability and sticking coefficient results are shown in Table 2. For comparison with previous studies (e.g., Veeraghattam et al. 2014 and Al-Halabi et al. 2007), we plot the sticking probability ($P(E_{\rm H},T_{\rm d})$) vs. the incident kinetic energies on 10 K and 70 K ice in Figures 1 and 3, respectively. In addition, in Figures 2 and 4, are plots of the sticking coefficient ($S(T_{\rm H},T_{\rm d})$) vs. the gas temperature for 10 K and 70 K ice, respectively.

\subsection{10 K Amorphous Water-Ice Slab}
Because, to our knowledge, there have been no studies which calculated sticking probabilities and coefficients on substrate temperatures other than 10 K and 70 K, we focus the majority of our discussion on those two slab temperatures. Figure 1 displays the sticking probability  plotted vs. the incident kinetic energy of the hydrogen atom. Virtually all the atoms ``stick" to the substrate at very low incident kinetic energies (i.e., 5 K and 10 K). Then, as the incident kinetic energy of the H atom increases, the sticking probability decreases. The probability decreases slowly at first and then at around $E_{\rm H} \approx 100$ K, the probability begins to decrease more rapidly. 

Our results (represented by dots) show good agreement with those results of Veeraghattam et al. (2014) (represented by triangles), in spite of the fact that that study used a TIP4P (rather than a TIP4P/Ice) potential for the water-water interaction and used a different sticking criterion (see Appendix). In addition, our results show strong agreement with those of Al-Halabi et al. (2002), who also utilized a TIP4P potential. With that being said, the results of Al-Halabi et al. (2002) were on a crystalline water-ice slab, rather than {an} amorphous one.

The results of Buch \& Zhang (1991) and those of Masuda et al. (1997) are not in good agreement with our study or those mentioned in the preceding paragraph. Buch \& Zhang (1991) obtain sticking probabilities that are much lower than those obtained here, whereas the sticking probabilities of Masuda et al. (1997) are in general higher. Buch \& Zhang (1991) used a small cluster of ice comprising of 115 water molecules. The cluster is highly irregular and as a result it is more difficult for the H atom to stick to the cluster \cite[]{alh02}. Masuda et al. (1997) performed simulations on a slab comprising of the same number of water molecules as our own (1000) but used a different potential to describe the water-water interaction, TIP2S \cite[]{mas97}. We cite the same discrepancy that Al-Halabi et al. (2002) reported to explain the difference between their results and those of Masuda et al. (1997). The study of Masuda et al. (1997) applied an incorrect expression for $Y_{0,0}$ that accounts for the greater chance of adsorption in their simulations \cite[]{alh02, vee14}. 

Figure 2 shows the sticking coefficient as a function of gas temperature. The sticking coefficient is very close to unity at low gas temperatures (5 and 10 K) and decreases as the temperature of the gas increases. Again, our results show strong agreement with those of Veeraghattam et al. (2014), despite the technical differences in our simulations. The small differences between our calculated coefficients could possibly be attributed to the different number of runs that we performed, or the difference in time step.\footnote{Veeraghattam et al. (2014) used a time-step of 0.5 fs, compared to 1.0 fs used here.} Regardless, all of the results of Veeraghattam et al. (2014) lie within the uncertainties of our own results. 

\subsection{70 K Amorphous Water-Ice}

Our results for the 70 K amorphous water-ice present a much more interesting case. These results largely disagree with those of previous studies. Figure 3, which displays the sticking probability as a function of incident kinetic energy,  illustrates the discrepancies. Our calculated sticking probabilities lie outside of the error bars of all three studies of comparison. Overall the trends are similar: in all three studies, the sticking probability decreases as the incident kinetic energy of the hydrogen atom increases. In contrast with the 10 K slab, the sticking probability for atomic hydrogen does not go to unity at low incident kinetic energy. Since the H$_2$O molecules of the 70 K slab have greater average rotational and intermolecular vibrational energy than those of the 10 K slab, correspondingly larger energy transfers when the slab and H-atom collide yield a reduced likelihood of sticking.   

The same reasoning is invoked to explain the differences between our results and those of Masuda et al. (1997) as for the 10 K amorphous water-ice. However, the discrepancies between our results and those of Veeraghattam et al. (2014) are more subtle. The use of a new potential, TIP4P/Ice instead of TIP4P, does not account for the discrepancy. We performed simulations using the TIP4P potential to test this and found that our results were still much higher than those of Veeraghattam et al. (2014). In fact, we found that the use of the new potential makes very little quantitative difference at all in these simulations. 

The only remaining difference between our results and those of Veeraghattam et al. (2014) is the criterion used for assessing sticking events. We defined a sticking event to be a run in which the H-atom had a final $z$-coordinate that was less than 13 \AA $\:$ \textit{and} had a final total energy less that -100 K. By contrast, Veeraghattam et al. (2014) defined a sticking event as a run in which the H-atom had a final $z$-coordinate that was less than 11 \AA $\:$ and did not consider the final total energy of the H-atom. These details make little difference on the 10 K ice, because at the end of the simulation atoms are very rarely still hopping around with high amplitudes. However, on 70 K ice it takes longer for atoms to give up their kinetic energy to the slab, because the water molecules in the slab have higher rotational and translational kinetic energies. As a result, an atom may still be bouncing on the surface of the substrate at the end of the run. We have selected a higher cutoff $z$ coordinate to include these bouncing atoms. We have also opted for a longer simulation time: 10 ps, compared to the 5 ps of Veeraghattam et al. (2014), in order to allow for sufficient relaxation on the warmer substrate. {We also performed a few trial simulations of 20 ps to confirm that 10 ps was an adequate selection. We found that the additional 10 ps had a negligible impact on the sticking or non-sticking status of the H-atom.} In addition, we have applied an energy cutoff of -100 K in the H-atom total energy as part of the sticking criterion. This was done to align with the sticking criterion used by by Al-Halabi et al. (2002). As a result of the contrasting definition with Veeraghattam et al. (2014), we observe large discrepancies between calculated probabilities and coefficients. However,  we do see that the sticking probability tends to be in closer agreement with the results of Veeraghattam et al. (2014) as the incident kinetic energy of the H-atom increases. This is because the ``bouncing atom" scenario play less of a role at higher incident kinetic energies, where the H atom generally either sticks, or leaves the simulation box within the simulation time.  

There are not many data points of Al-Halabi et al. (2002) to compare with and those that are present disagree with our predicted sticking probabilities. That fact that Al-Halabi et al. (2002) performed their simulations on crystalline ice is the most likely explanation for the discrepancy, since all other features of the calculations (potentials, sticking criteria, etc.) are either similar or the same.

Because the sticking coefficient is a function of the sticking probability, our calculated sticking coefficients are much larger than those of Veeraghattam et al. (2014), reflecting the discrepancy in our sticking probabilities. This is shown in Figure 4, where the sticking coefficient is plotted against the gas temperature. Again there is better agreement at higher gas temperatures, where the particular criterion of a sticking event has less of an effect. 

\subsection{Other Amorphous Water-Ice Slab Temperatures}  
Having discussed the results on the low-end and high-end of the slab temperatures studied, it is reasonable to assume that the sticking probabilities and coefficients for the in-between slab temperatures lie somewhere in-between the above results. Table 2 confirms that that is indeed the case. We provide these results for astrophysical modeling purposes. As the slab temperature increases, in general, the sticking probability and coefficient decrease for each respective H-atom incident kinetic energy. 

Figure 5 shows the sticking coefficient for each gas temperature (5 K - 400 K) as a function of substrate temperature. This plot demonstrates some key features of these simulations. For all but the highest gas temperatures, the sticking coefficient decreases with increasing substrate temperature, as one would expect. However, the overall drop in sticking coefficient from the value at substrates temperature of 10 K to 70 K decreases as gas temperature increases. By gas temperatures of 300 K and 400 K,\footnote{Represented by crosses and diamonds respectively.} the sticking coefficient is very nearly constant in substrate temperature, at least over the range of temperatures studied. This behavior can be understood by considering the average kinetic energy of the gas atoms. {At high gas temperature the average kinetic energy is so high compared to that of the substrate H$_2$O molecules that energy transfers\footnote{{i.e. the thermalization of the H-atoms with the substrate}} to and from the substrate are insensitive to substrate temperature {\cite[]{aba05}}.} 

\subsection{Astrophysical Implications}
{As shown previously by \cite{vee14}, the sticking coefficients
for 10 K ice obtained in our MD simulations gives results 
significantly larger than those predicted by \cite{hol70}. The latter is adopted in most astrochemical models involving ice-surface chemistry. Since the formation of H$_2$ on the ice surface is linearly proportional to the sticking coefficient, the adoption of the current values would be expected to result in both an increased efficiency in the molecular hydrogen formed on the ice surface as well as faster removal of atomic hydrogen from the gas phase. Further, since the current sticking coefficients on 70 K ice can be as much as 30\% larger than those calculated earlier in \cite{vee14}, we can expect the  H$_2$ efficiency to be relatively enhanced even as the surface temperature increases (see Fig. 5) until water begins to evaporate from the surface above $\sim$100~K. Such an effect was found for the sticking of H$_2$ and other molecular species on amorphous ice in the time-dependent, hot core warm-up modeling of \cite{he16}.}

\section{Conclusions}

We have reported the sticking probabilities and coefficients for nine gas temperatures of atomic hydrogen (10 - 400 K) across seven different substrate temperatures (10 K - 70 K). We present our results so that they may be used in future astrophysical modeling simulations. The sticking of hydrogen atoms on amorphous water-ice is of significant importance in the ISM, and its understanding is necessary before more complicated molecular species, which form through grain-mediated reactions, can be understood. This study shows in detail how grain temperature affects sticking probabilities and coefficients, thereby showing how grain temperature affects the rate of H$_2$ formation. 

It is important however to point out some of the limitations of the study. First, the nature of the simulation (i.e., launching a single H atom at a surface and determining its adsorption) is different from the actual adsorption process in the ISM, in which a dilute cloud of gas comprising of H atoms passes through these dusty regions. In addition, the lack of a fully quantum treatment may be cause for some concern, because quantum effects tend to be more important for lighter atoms. 

This study has shown that the criterion for assessing a sticking event can greatly effect sticking probabilities. In the appendix of the paper we discuss the methods by which we have arrived at our sticking criterion. We have explored many different criteria (e.g. velocity, standard deviation of position, etc.) for this process and have determined that our sticking criterion is the most appropriate choice for wide application. Further studies involving different species which have previous data (e.g. H$_2$ on H$_2$O ice) could confirm our criterion for sticking.

The authors would like to thank Vijay Veeraghattam for his useful comments and guidance throughout this entire project. {We also would like to thank the referee whose comments greatly improved the final version of the manuscript.}

\newpage
\section{Appendix: How to Define a Sticking Event}

The purpose of this section is to provide physical reasoning for our definition of a sticking event. A sticking event occurs when a hydrogen atom has adsorbed to the water-ice substrate. There are three main outcomes of these simulations: back scattering events, trapping-desorption events, and adsorption events \cite[]{alh04}. A back scattering or elastic event is one in which the atom leaves the simulation box before the simulation ends. This event is a clear non-sticking event. An adsorption event is one in which the atom very clearly sticks to the surface within the timescale of the simulation. {This is determined by examining a) the final position of the hydrogen atom and b) the final total energy of the atom. Clear sticking events have a final position that is either embedded in the substrate or very close to the top of the substrate and low total energy, indicating that the atom has given up most of its kinetic energy and is bound to the surface.} Trapping-desorption events, however, are difficult to define within classical MD. This is a case where an atom appears to be adsorbed to the substrate, but then gains kinetic energy from the substrate which causes the atom to desorb. The uncertainty with regards to the desorption is that it may occur after the timescale of the MD simulation. {Trapping-desorption events have time scales on the scale of \textit{nanoseconds} whereas our simulation has a timescale of 10 \textit{picoseconds} \cite[]{alh04}.} In the instance of trapping-desorption events, an atom may be thermally hopping with large amplitude at the end of the run, yet the atom will still be confined to the simulation box. In classical MD simulations these trapping desorption runs must be designated as a sticking or non-sticking event. This designation is dependent on the particular definition of adsorption.     

  It is worth briefly considering the definitions of adsorption in previous studies of this nature.  Buch \& Zhang (1991), defines a sticking trajectory as one that has a final energy that is less than -100 K. They also state that no questionable runs occurred in their simulations. Either the atom stuck, or it was scattered. K. Masuda et al. 1997 studies the sticking probability of hydrogen atoms on the icy mantle of dust grains. They define a sticking event as one where an ``incident H atom resided on the surface of the amorphous water ice after 5000 fs of simulation time." Whereas a scattering/non-sticking event was one in which the atom returned to its initial height at the start of the simulation \cite[]{mas97}. This definition neglects the possibility of trapping-desorption trajectories and gives no quantitative value for what was considered to be the cutoff for a sticking event. 
  
  Al-Halabi et al. (2004) who studied CO adsorption on amorphous water ice, defines a sticking event as one that exhibits more than one turning point in the $z$ coordinate such that 1) the final $z$ coordinate of position $Z_{\rm f}$ $ \le 12.5 $ \AA $\;$ and 2) the final energy of the adsorbed CO molecule is below $-k_B T_{\rm d}$, where $T_{\rm d}$ is the temperature of the slab, and $k_B$ is the Boltzmann constant. {The values for $Z_{\rm f}$ are converted from the original number to allow comparison to our simulation box, in which the top of the substrate has an average height of 10 \AA.} Al-Halabi et al. (2002) employed a similar definition, but does without the $Z_{\rm f}$ cutoff and changed the final total energy cutoff to be -100 K. Finally, Veeraghattam et al. (2014) says ``a sticking event occurs when the H atom stops bouncing on the surface and remains stuck for the rest of the simulation." However, we found that the H atom does not stop bouncing on the surface within the simulation time. The H atom, even if embedded in the icy substrate continues to bounce with small amplitudes on or within the substrate. Upon further investigation, Veeraghattam et al. (2014) applied a restrictive height cutoff value ($Z_f = 11$ \AA). 
  
{We initially took a sticking event to be any such run that leaves the simulation box before the timescale of the simulation. In doing so, we noticed a discrepancy between our results and those of Veeraghattam et al. (2014).} Figures 6 and 7, which illustrate the reason for this discrepancy, show the sticking probability as a function of cutoff value $Z_{\rm f}$ for 10 and 70 K substrates, respectively. We let $Z_{\rm f}$ equal 100 values between 0.0 and 1.0 {in reduced units, where 1.0 corresponds to 20 \AA $\:$ and 0.5 corresponds to 10 \AA.} We then compute the sticking probability at each $Z_{\rm f}$.  From Figure 6, we can see that on a 10 K substrate the sticking probabilities plateau fairly quickly. The approximate top of the substrate is at 0.5 (10 \AA) in reduced units. Probabilities plateau around 0.55 - 0.6, which is about 11-12 \AA. In addition, in the region of 0.55-0.6 the probabilities are not varying significantly. {Although uncertainties are not plotted here for aesthetic reasons, the probabilities in this region would be within uncertainties.}  Compare this to Figure 7, in which probabilities plateau around 0.65 (13 \AA), and the probabilities are very different from those calculated at 0.55 (11 \AA). Figure 7 further demonstrates the reason for our higher sticking probabilities when compared to those results of Veeraghattam et al. (2014) on 70 K ice, while Figure 6 demonstrates the reason that our results agree on 10 K ice. As a result of Figure 7, we decided to select $Z_f = 0.65$ as our cutoff value in all sticking calculations. 

However, based on the studies of Al-Halabi et al. (2002, 2004, and 2007), we decided apply a second criterion for sticking. After exploring other parameters such as the final mean $z$ velocity of the H atom in the last 0.3 ps of a run and the standard deviation in $z$ position of the H atom in the final 0.3 ps of a run, we decided that final total energy ($E_{\rm f}$) was the appropriate choice. Figures 8 and 9 plot the final total energy value against the final $z$ coordinate value for each of the 300 runs of a 150 K atom on 10 and 70 K ice, respectively. Atoms in the dense region towards the upper right of each plot are clear non-sticking events. These are atoms that have left the simulation box within the timescale of the simulation. Atoms whose location is near to the bottom left corner of the plots are definite sticking events. These are atoms that are embedded in the slab with virtually zero kinetic energy. 

Comparing the middle regions of Figure 8 to Figure 9, one is able to notice some significant differences. For instance, in Figure 8, the atoms on the 10 K substrate that have not left the simulation box (i.e. those atoms not in the upper right hand corner) are relatively densely packed when compared to those atoms on the 70 K substrate. This shows that the higher temperature of the substrate is affecting where the atoms will be and what energy the atoms will have at the end of the simulation. Note also that on the 10 K substrate the atoms seem to be grouped together in clumps, whereas there is no grouping behavior on the 70 K substrate. {The reasons for this clumping behavior are not completely clear at the present but they are likely due to regions of our slab configuration which are energetically favorable for the H-atoms to settle. There appear to be ``pockets" in our slab that the atoms have a preference to congregate towards. By contrast, in the 70 K substrate, there are no energetically favorable sites for the atoms to settle into so all regions of Z$_{\rm f}$ and final total energy are equally likely. To summarize, the clumping behavior is likely due to the differences in the particular amorphous ice slab configuration between the 10 K and 70 K slabs. Further analysis would be necessary to confirm this claim.}

Based on Figures 8 and 9 we can draw a box around the atoms which have stuck to the substrate using $Z_{\rm f}$ and $E_{\rm f}$. The atoms in the box have adsorbed to the substrate whereas those atoms which are outside the box have not adsorbed to the substrate. Figure 10 displays how the selection of $E_{\rm f}$ with a fixed $Z_{\rm f} = 0.65$ affects sticking probability on 70 K ice. Note that we only show the 70 K substrate here because it is the 70 K substrate that we are using to choose our sticking event parameters. We see that the sticking probability plateaus around -500 K, but in order to be conservative about our selection and to agree with the parameter of Al-Halabi et al. (2002), we select $E_{\rm f} = -100 $ K.

\begin{figure}
\centering
\plotone{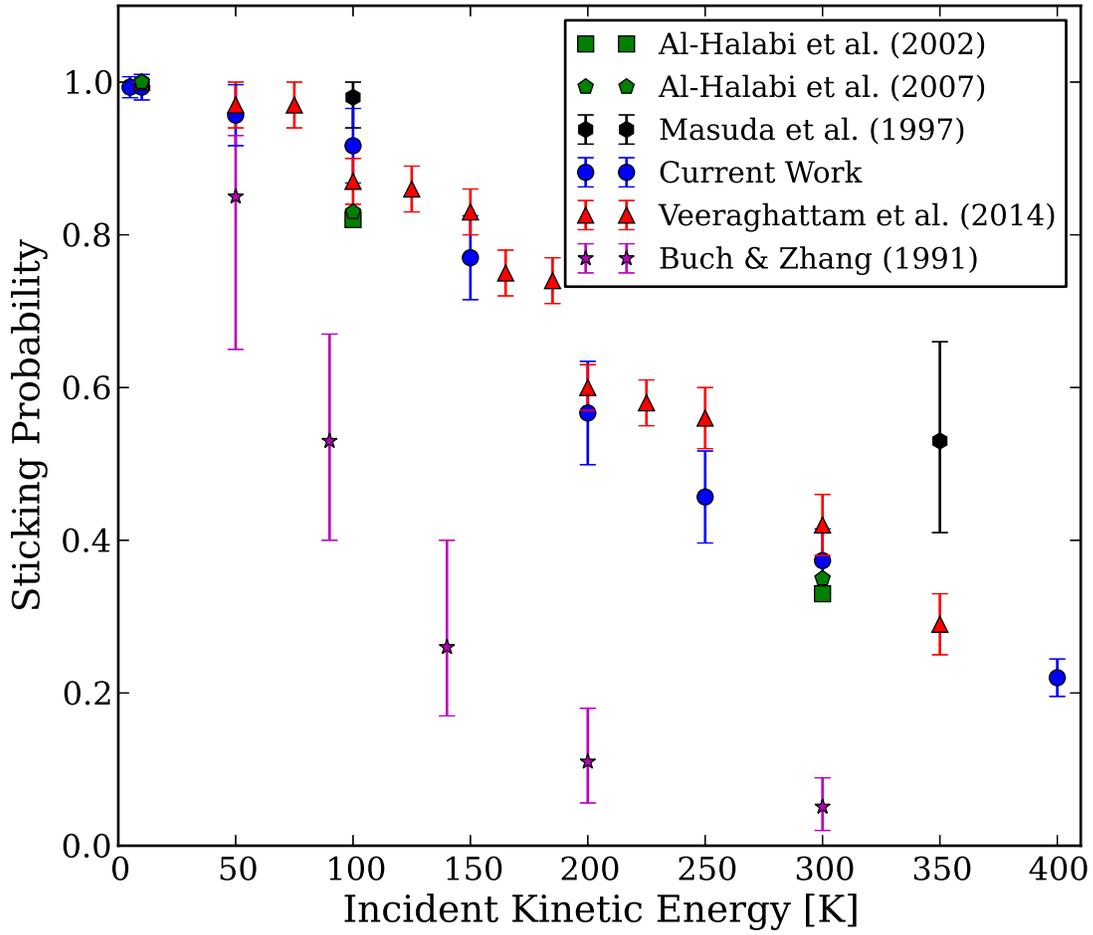}
\caption{Sticking probability on 10 K amorphous water ice. Note that the results of Al-Halabi et al. (2002) are on crystalline ice. }
\end{figure}

\begin{figure}
\centering
\plotone{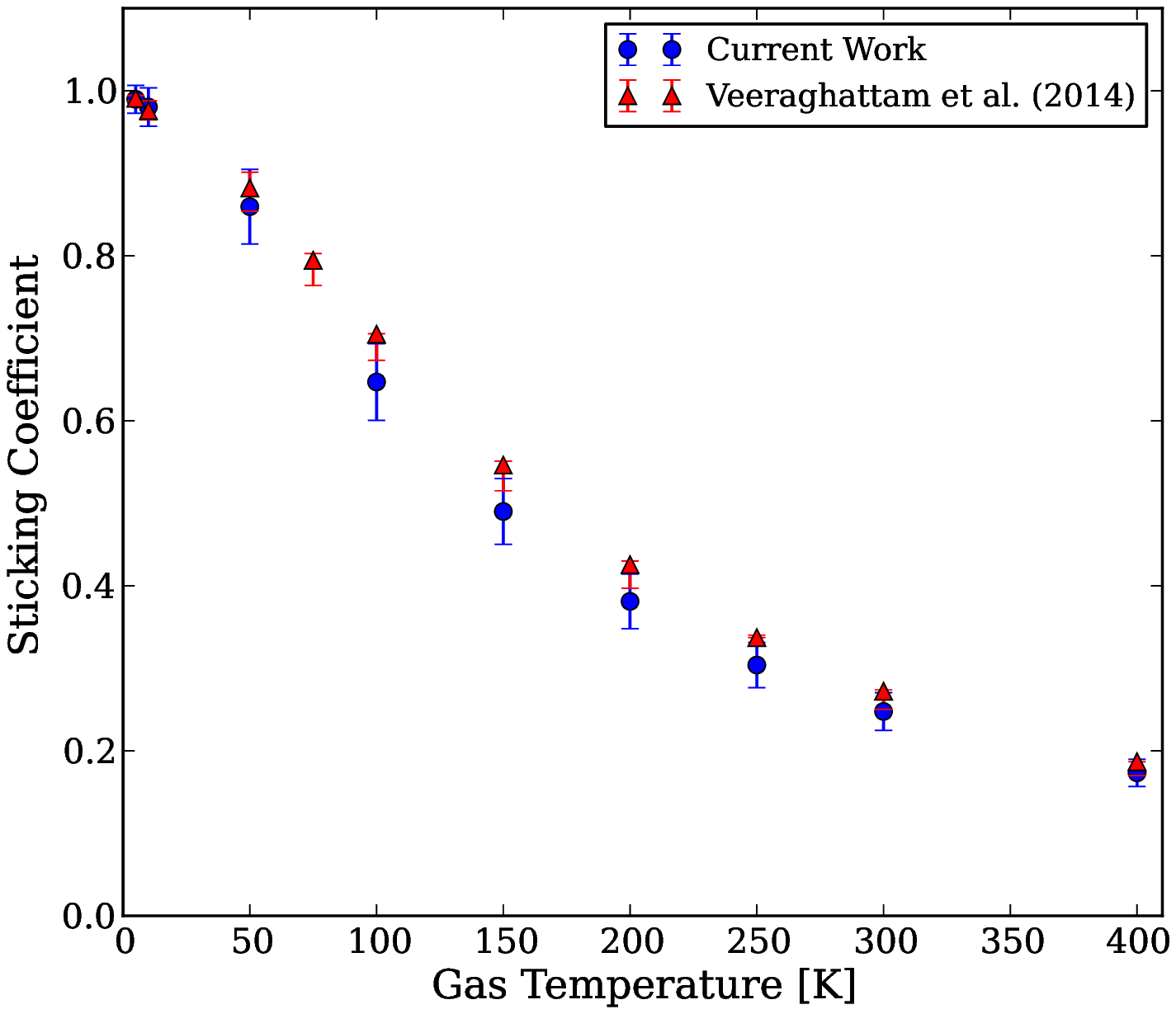}
\caption{Sticking coefficient on 10 K amorphous water ice. }
\end{figure}

\begin{figure}
\centering
\plotone{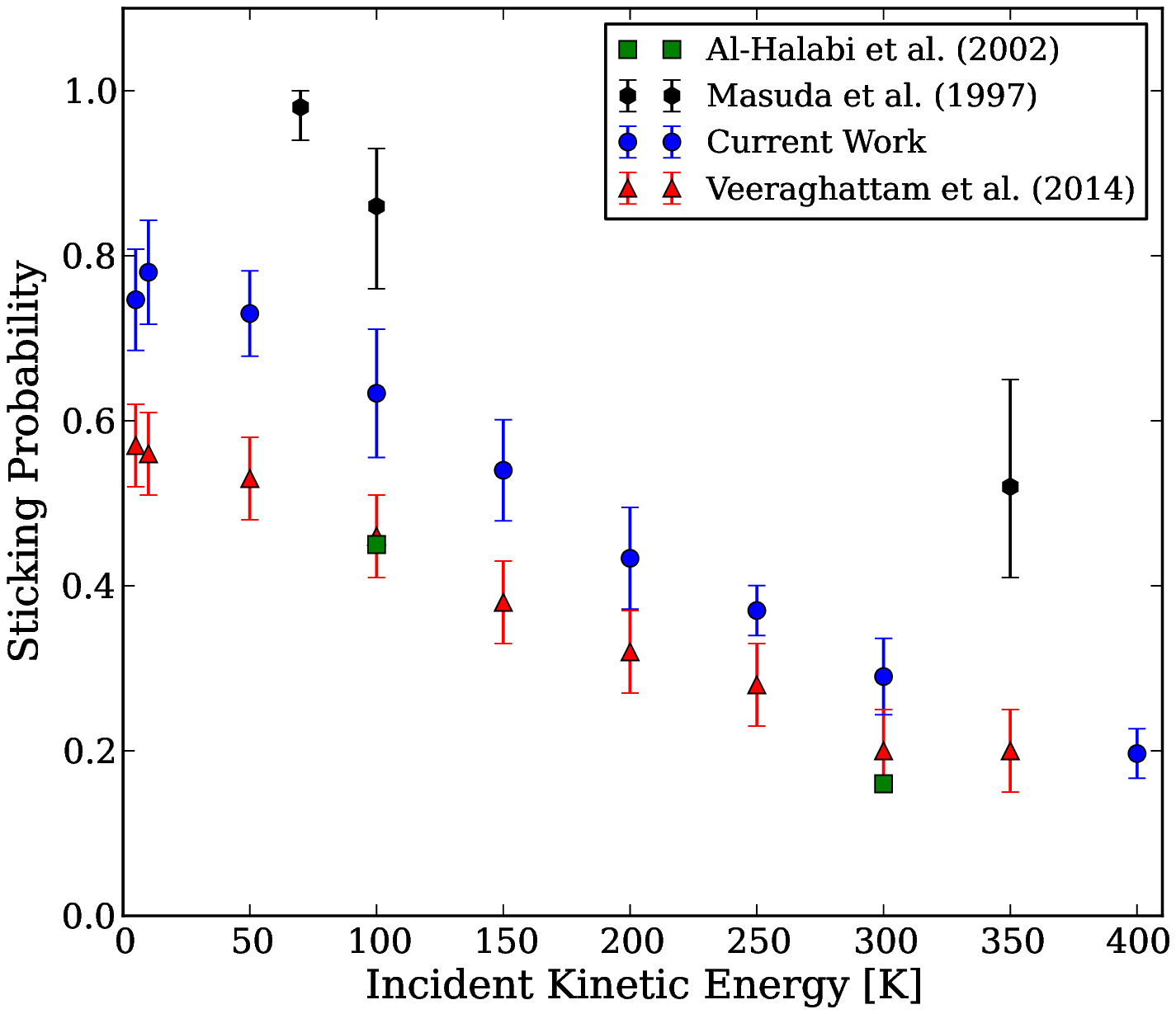}
\caption{Sticking probability on 70 K amorphous water ice. Note that the results of Al-Halabi et al. (2002) are on crystalline ice. }
\end{figure}

\begin{figure}
\centering
\plotone{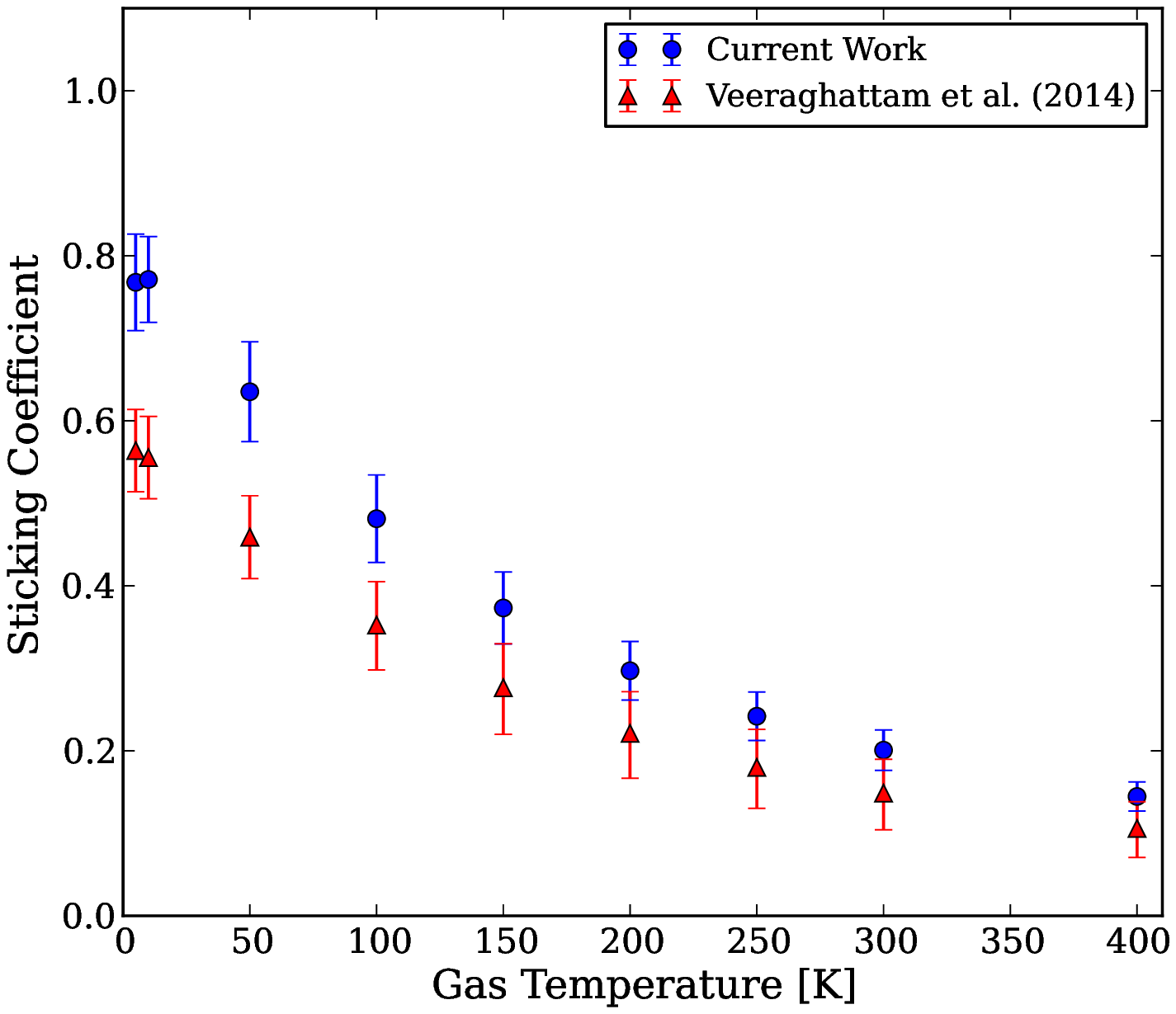}
\caption{Sticking coefficient on 70 K amorphous water ice. }
\end{figure}

\begin{figure}
\centering
\plotone{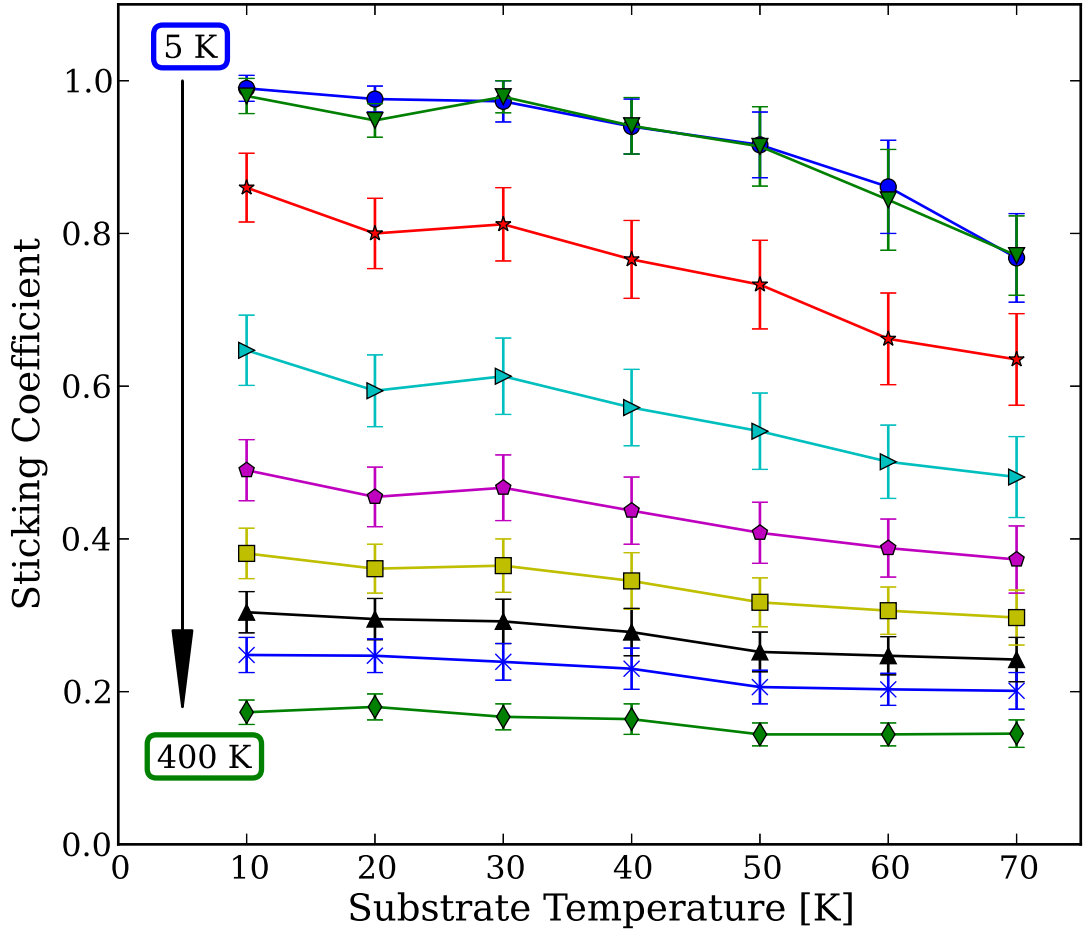}
\caption{Sticking coefficient as a function of substrate temperature for each specific gas temperature. The key is as follows: Circles: 5 K, Downward Triangles: 10 K, Stars: 50 K, Right Triangles: 100 K, Pentagons: 150 K, Squares: 200 K, Triangles: 250 K, Crosses: 300 K, Diamonds: 400 K.     }
\end{figure}

\begin{figure}
\centering
\plotone{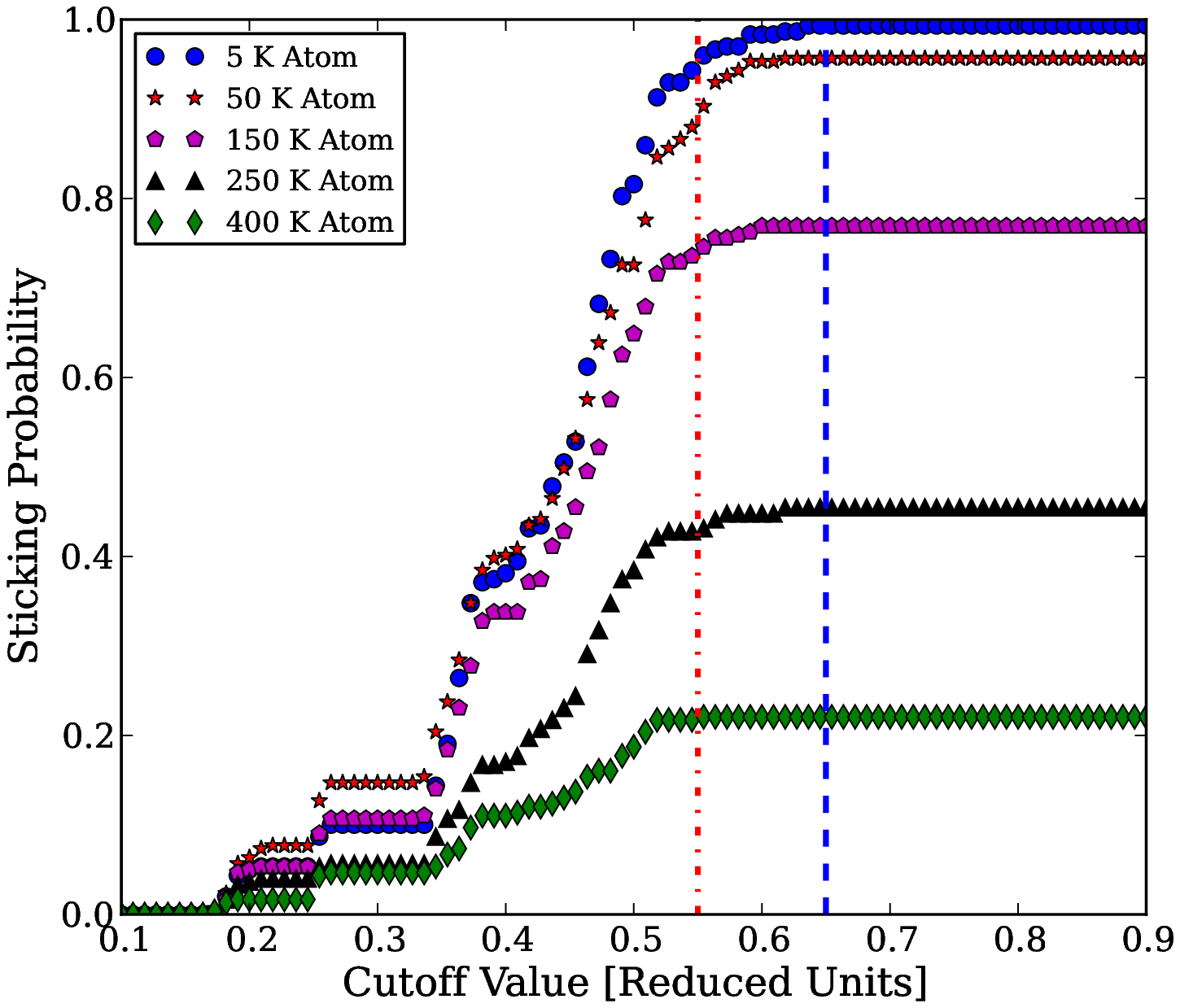}
\caption{Sticking probability as a function of $Z_{\rm f}$ cutoff value for the 10 K substrate. The dot-dashed line represents the cutoff value of Veeraghattam et al. (2014) whereas the dashed line represents the cutoff value of the current study. }
\end{figure}

\begin{figure}
\centering
\plotone{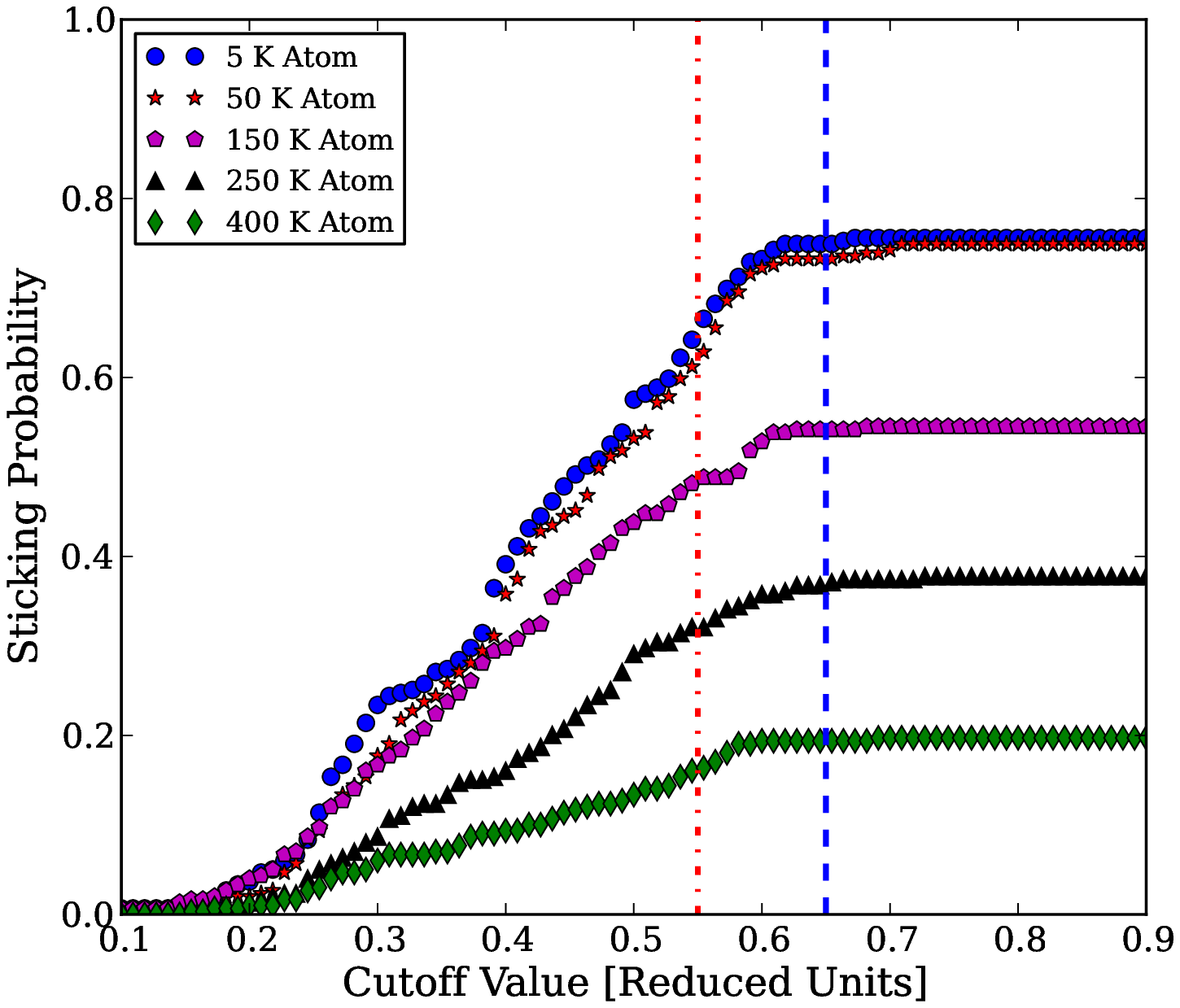}
\caption{Sticking probability as a function of $Z_{\rm f}$ cutoff value for the 70 K substrate. The dot-dashed line represents the cutoff value of Veeraghattam et al. (2014) whereas the dashed line represents the cutoff value of the current study. }
\end{figure}

\begin{figure}
\centering
\plotone{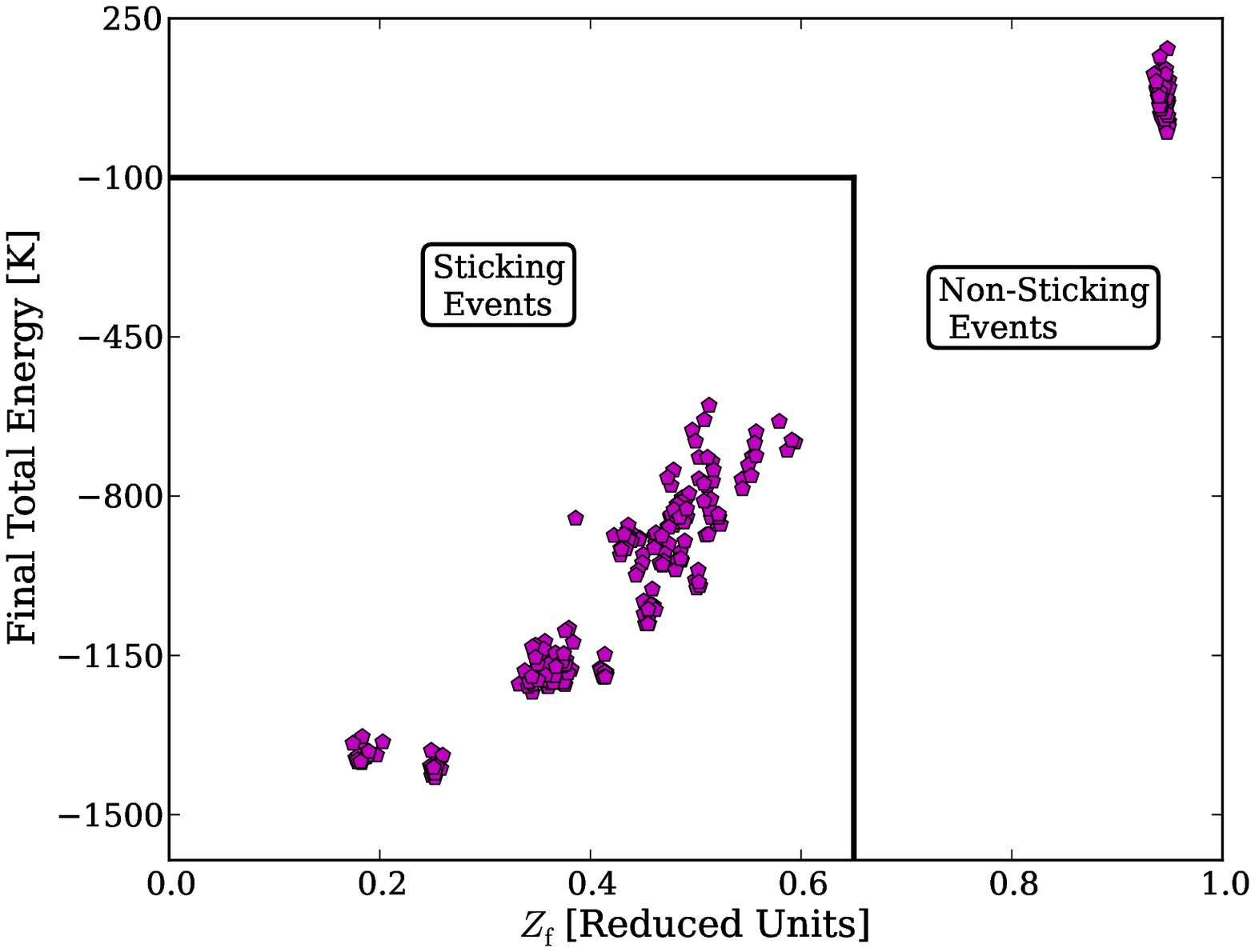}
\caption{ Final total energy value plotted against final $z$ coordinate value (reduced units) for 300 trajectories of 150 K H atoms on 10 K ice. }
\end{figure}

\begin{figure}
\centering
\plotone{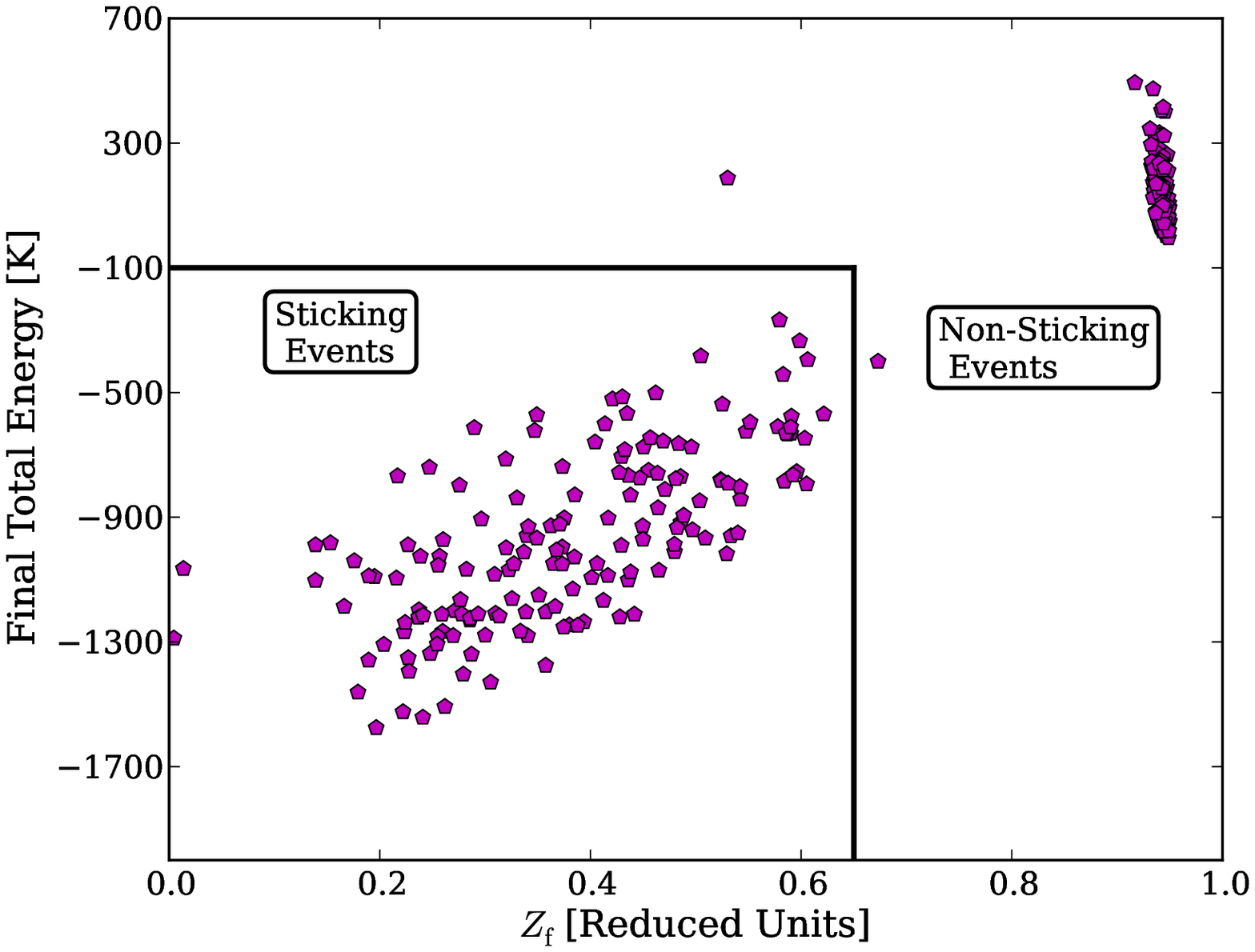}
\caption{ Final total energy value plotted against final $z$ coordinate value (reduced units) for 300 trajectories of 150 K H atoms on 70 K ice. }
\end{figure}

\begin{figure}
\centering
\plotone{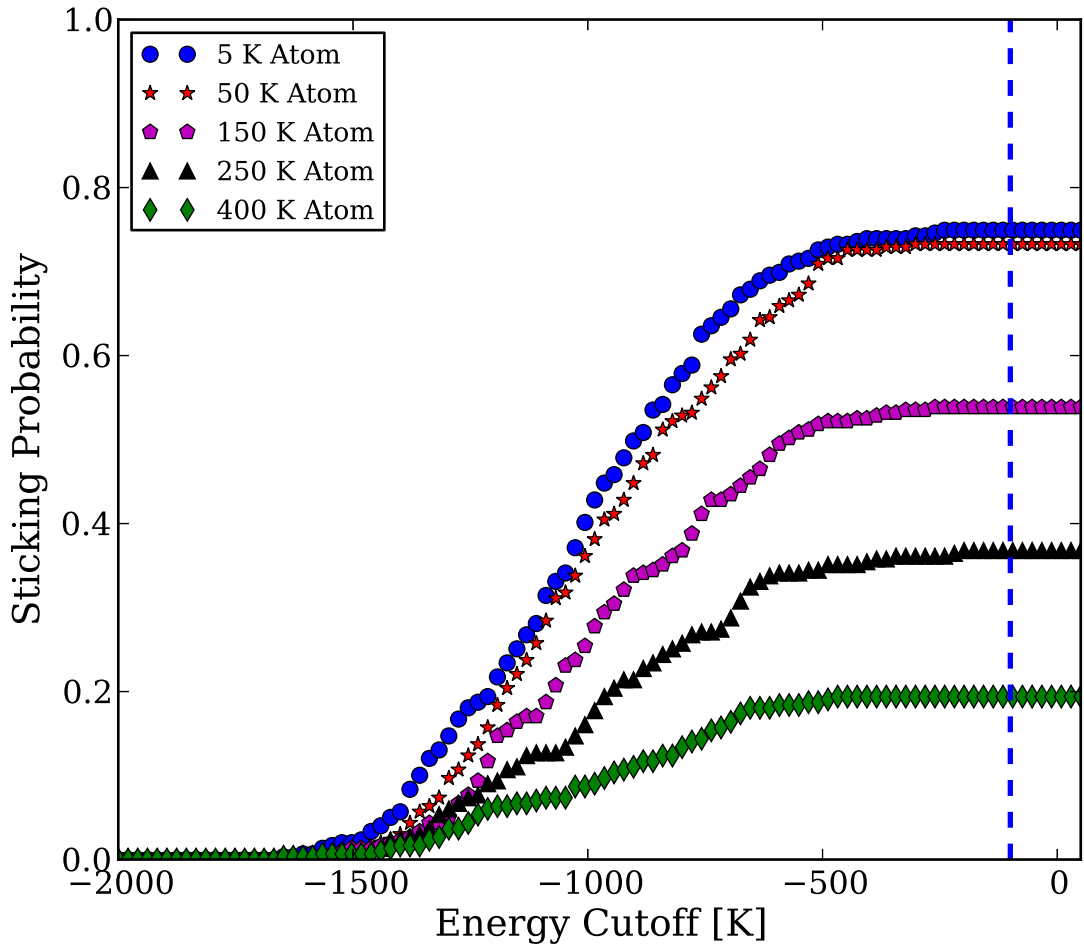}
\caption{Sticking probability as a function of $E_{\rm f}$ cutoff value with a fixed $Z_{\rm f}$ cutoff value of 0.65 for the 70 K substrate. }
\end{figure}

\begin{table}
\begin{center}
\caption{Parameters for the H-H$_2$O Interaction Potential \cite[]{zha91}.
\vspace{0.5cm}
\label{tbl-1}}
\begin{tabular}{c c c}

\tableline\tableline
$l,m$ \hspace{1.5cm} & $\epsilon_{l,m}$ \hspace{1.5cm} & $\sigma_{l,m}$ \\ 
  \hspace{1.5cm}    & [K] \hspace{1.5cm} & [\AA] \\ \tableline
0,0    \hspace{1.5cm}    & 222.1  \hspace{1.5cm}   & 3.33 \\ 
1,0    \hspace{1.5cm}    & -19.6  \hspace{1.5cm}    & 3.00 \\
2,0    \hspace{1.5cm}    & 17.7   \hspace{1.5cm}    & 2.98 \\
2,2    \hspace{1.5cm}    & 73.3   \hspace{1.5cm}    & 2.92 \\ \tableline  
\end{tabular}
\end{center}
\end{table}  

\begin{deluxetable}{c c c c c c}
\tablewidth{0pt}
\tablecaption{Predicted sticking Probabilities and Coefficients for atomic hydrogen on amorphous water-ice. $T_{\rm d}$ is the ice-substrate temperature, $E_{\rm H}$ is the incident kinetic energy of the hydrogen atoms, and $T_{\rm H}$ is the temperature of the atomic hydrogen gas. $\sigma_{\rm P}$ and $\sigma_{\rm S}$ represent the uncertainty in the sticking probability ($P(E_{\rm H},T_{\rm d})$) and the uncertainty in the sticking coefficient ($S(T_{\rm H},T_{\rm d})$), respectively.}
\tablehead{
\colhead{$T_{\rm d}$ [K]}           & \colhead{$ E_{\rm H},T_{\rm H} $ [K]}      &
\colhead{$P(E_{\rm H},T_{\rm d})$}          & \colhead{ $\sigma_{\rm P}$}  &
\colhead{$S(T_{\rm H},T_{\rm d})$}          & \colhead{$\sigma_{\rm S}$} }
\startdata
10 & 5 & 0.993 & 0.014 & 0.990 & 0.017 \\ 
   & 10 & 0.993 & 0.017 & 0.980 & 0.023 \\ 
   & 50 & 0.957 & 0.040 & 0.860 & 0.045 \\ 
   & 100 & 0.917 & 0.049 & 0.647 & 0.046 \\ 
   & 150 & 0.770 & 0.055 & 0.490 & 0.040 \\ 
   & 200 & 0.567 & 0.068 & 0.381 & 0.033 \\ 
   & 250 & 0.457 & 0.060 & 0.304 & 0.027 \\ 
   & 300 & 0.373 & 0.042 & 0.248 & 0.023 \\ 
   & 400 & 0.220 & 0.024 & 0.173 & 0.016 \\ \hline
20   & 5 & 1.000 & 0.000 & 0.976 & 0.017 \\  
   & 10 & 0.977 & 0.030 & 0.948 & 0.022 \\ 
   & 50 & 0.917 & 0.027 & 0.800 & 0.046 \\ 
   & 100 & 0.843 & 0.064 & 0.594 & 0.047 \\ 
   & 150 & 0.640 & 0.059 & 0.455 & 0.039 \\ 
   & 200 & 0.553 & 0.066 & 0.361 & 0.032 \\ 
   & 250 & 0.410 & 0.056 & 0.295 & 0.027 \\ 
   & 300 & 0.297 & 0.037 & 0.247 & 0.022 \\ 
   & 400 & 0.223 & 0.021 & 0.180 & 0.017 \\  \hline
30   & 5 & 0.963 & 0.031 & 0.973 & 0.027 \\ 
   & 10 & 0.967 & 0.027 & 0.979 & 0.021 \\ 
   & 50 & 0.923 & 0.041 & 0.812 & 0.048 \\ 
   & 100 & 0.800 & 0.062 & 0.613 & 0.050 \\ 
   & 150 & 0.770 & 0.051 & 0.467 & 0.043 \\ 
   & 200 & 0.523 & 0.059 & 0.365 & 0.035 \\ 
   & 250 & 0.440 & 0.054 & 0.292 & 0.029 \\ 
   & 300 & 0.377 & 0.064 & 0.239 & 0.024 \\ 
   & 400 & 0.220 & 0.033 & 0.167 & 0.017 \\  \hline
40   & 5 & 0.937 & 0.039 & 0.940 & 0.036 \\ 
   & 10 & 0.937 & 0.034 & 0.941 & 0.037 \\ 
   & 50 & 0.900 & 0.046 & 0.766 & 0.051 \\ 
   & 100 & 0.743 & 0.057 & 0.572 & 0.050 \\ 
   & 150 & 0.647 & 0.063 & 0.437 & 0.044 \\ 
   & 200 & 0.547 & 0.057 & 0.345 & 0.037 \\ 
   & 250 & 0.420 & 0.052 & 0.278 & 0.031 \\ 
   & 300 & 0.310 & 0.047 & 0.230 & 0.027 \\ 
   & 400 & 0.210 & 0.032 & 0.164 & 0.020 \\ \hline
50   & 5 & 0.917 & 0.033 & 0.916 & 0.043 \\ 
   & 10 & 0.910 & 0.047 & 0.914 & 0.052 \\ 
   & 50 & 0.870 & 0.064 & 0.733 & 0.058 \\ 
   & 100 & 0.707 & 0.057 & 0.541 & 0.050 \\ 
   & 150 & 0.610 & 0.054 & 0.408 & 0.040 \\ 
   & 200 & 0.510 & 0.064 & 0.317 & 0.032 \\ 
   & 250 & 0.387 & 0.044 & 0.252 & 0.026 \\ 
   & 300 & 0.307 & 0.040 & 0.206 & 0.022 \\ 
   & 400 & 0.180 & 0.023 & 0.144 & 0.015 \\ \hline
60   & 5 & 0.853 & 0.055 & 0.861 & 0.061 \\ 
   & 10 & 0.873 & 0.062 & 0.844 & 0.066 \\ 
   & 50 & 0.767 & 0.069 & 0.662 & 0.060 \\ 
   & 100 & 0.647 & 0.058 & 0.501 & 0.048 \\ 
   & 150 & 0.530 & 0.059 & 0.388 & 0.038 \\ 
   & 200 & 0.457 & 0.047 & 0.306 & 0.031 \\ 
   & 250 & 0.410 & 0.034 & 0.247 & 0.025 \\ 
   & 300 & 0.340 & 0.033 & 0.203 & 0.021 \\ 
   & 400 & 0.200 & 0.021 & 0.144 & 0.015 \\  \hline
70   & 5 & 0.747 & 0.061 & 0.768 & 0.058 \\ 
   & 10 & 0.780 & 0.063 & 0.771 & 0.052 \\ 
   & 50 & 0.730 & 0.052 & 0.635 & 0.060 \\ 
   & 100 & 0.633 & 0.078 & 0.481 & 0.053 \\ 
   & 150 & 0.540 & 0.061 & 0.373 & 0.044 \\ 
   & 200 & 0.433 & 0.062 & 0.297 & 0.036 \\ 
   & 250 & 0.370 & 0.030 & 0.242 & 0.029 \\ 
   & 300 & 0.290 & 0.046 & 0.201 & 0.024 \\ 
   & 400 & 0.197 & 0.030 & 0.145 & 0.018
 \enddata
\end{deluxetable}

\end{document}